\documentclass[11pt]{article}
\usepackage{latexsym}
\usepackage{epsfig,amssymb,euscript}
\usepackage{amsmath}

\def\be{\begin{equation}}
\def\ee{\end{equation}}
\def\bea{\begin{eqnarray}}
\def\eea{\end{eqnarray}}

\input epsf
\begin{document}
{\small
\hfill LPTHE-P05-06
\par\hfill UB-ECM-PF-05/30
\par\hfill hep-th/0512109
}
\begin{center}
{\LARGE \bf New results for AdS/CFT and beyond}
\end{center}
%
%
\begin{center}
{\bf M. Bertolini$\,^a$, F. Bigazzi$\,^{b,c}$, A. L. Cotrone$\,^{d}$}
\end{center}
\begin{center}
{\small 
\textit{a} SISSA/ISAS and INFN, Via Beirut 2; I-34014 Trieste, Italy.\\
\textit{b} LPTHE, Universit\'es Paris VI and VII, 4 place Jussieu; 75005, Paris, France.\\
\textit{c}  INFN, Piazza dei Caprettari, 70; I-00186  Roma, Italy.\\
\textit{d} Departament ECM, Facultat de F\'isica, Universitat de Barcelona and \\Institut
de Fisica d'Altes Energies, Diagonal 647, E-08028 Barcelona, Spain.\\

{\small\tt bertmat@sissa.it, bigazzi@lpthe.jussieu.fr, cotrone@ecm.ub.es}
}
\end{center}
%
%
%
{\small \noindent
We report on some recent results within the string/gauge theory correspondence, both in the conformal and in
the non conformal cases, for a recently found class of ${\cal N}=1$ dual pairs. These results provide the
first cross check of AdS/CFT and field theory techniques
like $a$-maximization. Moreover, they furnish new examples of cascading gauge theories and the first instance of 4d dynamical supersymmetry breaking embedded in the correspondence.}
\vspace{15pt}
%
\section{Introduction}
Placing branes at singularities is a useful tool to extend the string/gauge theory correspondence beyond its original ${\cal N}=4$ realization \cite{malda}. If the singularity is an orbifold, the low energy dynamics on the branes can be deduced from known techniques, but
 for more generic singularities this is not an easy task. 

During the last months some progress has been made in the field. General rules have been given for the case where the branes are placed at the tip of a toric non compact Calabi-Yau cone. A six dimensional (6d) CY cone has a base which is a 5d Sasaki-Einstein manifold $X_5$. The low energy effective dynamics of a stack of $N$ D3-branes at the tip of a 6d toric CY cone is described by a quiver four dimensional ${\cal N}=1$ superconformal field theory (SCFT). The moduli space of the SCFT has, by construction, a mesonic branch corresponding to copies of the singular CY cone. Precise informations on the gauge group, matter content and superpotential of the SCFT have been recently found to be encoded by the toric geometrical properties of the singularity through dimer models \cite{dimers} whose physical relevance was recently explained in \cite{vafa}. 

If we know the SCFT and the details of the geometry, we can explicitly check the AdS/CFT correspondence. A SCFT engineered as above is equivalent, through the correspondence, to Type IIB string on a supergravity background with $AdS_5\times X_5$ metric, $N$ units of $F_5$ flux through $X_5$ and constant dilaton and axion. When the background has large radius of curvature and the string coupling is small, the string model boils down to its low energy effective limit, which is classical IIB supergravity on the above background. On the field theory side, the limit corresponds to a large $N$, large 't Hooft coupling regime, which can be thus remarkably captured by the dual classical supergravity description. Fields on AdS are mapped to SCFT operators whose correlation functions can be evaluated from the classical dual description.




One could wonder how any quantitative check
of the gauge/gravity duality can actually be done. This is indeed possible  
thanks to supersymmetry and conformal invariance. There are field theory
observables, like some anomaly charges or the R-charges of the matter
fields, which can be exactly given, at the conformal IR fixed point,
provided the global symmetries at that point are exactly known. For a
generic field theory this is indeed a difficult task. But for SCFT's having a
supergravity dual, the global symmetries are
immediately given by construction: they are inherited, via AdS/CFT,
from the symmetries of the dual background. 

Until recently, explicit checks of the above correspondence were given only for $X_5$ being $S^5$ (in which case the ``cone'' is simply ${\mathbb C}^3$) or $T^{1,1}$ (the base of the conifold). The corresponding SCFT's are ${\cal N}=4$ $SU(N)$ SYM and an ${\cal N}=1$ $SU(N)\times SU(N)$ quiver theory \cite{kw} respectively. This scenario was enriched thanks to a newly found family of AdS/CFT ${\cal N}=1$
dual pairs, characterized by a new series of SE manifolds, dubbed $Y^{p,q}$, whose metric was explicitly given in \cite{gmswSE}. The technical progresses referred to as above took a crucial input from the study of the correspondence for these new cases. 

The $Y^{p,q}$ manifolds  ($p>q$ coprime integers) have an $SU(2)\times U(1)^2$ isometry group. Moreover they  have $S^2\times S^3$ topology, just as $T^{1,1}$. This has at least two interesting consequences. First, the global symmetry group of the dual SCFT's will result to be $SU(2)\times U(1)\times U(1)\times U(1)_R$. The additional abelian factor with respect to the above isometry group is related to the Kaluza-Klein reduction of $C_4$ on the topological $S^3$. Second, one can consider the addition of D5-branes, wrapped on $S^2$, to the stack of D3's at the singularity. The wrapped D5's behave as fractional D3's and the low energy effective dynamics of the whole system of branes is described by a non conformal gauge theory with logarithmic running couplings \cite{kn}. Studying the correspondence in this case reveals to be of evident interest.
 
While ${\cal N}=4$ SYM is a finite theory, conformal at any value of the coupling, a
common feature of the conifold theory and the SCFT's in the new class  is that they are quivers 
undergoing a non-trivial RG-flow whose end is an interacting IR fixed point. There are however two key properties of the duals of $Y^{p,q}$ which are not shared with the conifold theory. First, the R-charges at the IR fixed point cannot be
completely fixed using just (super)symmetries and ABJ anomaly-free arguments. To fix them one needs to invoke
new field theory tools, in particular the $a$-maximization prescription \cite{iw03}. Thus, checking the gauge/gravity duality in this case amounts to cross checking AdS/CFT and $a$-maximization. Second, the non conformal versions of the duality, though displaying an RG-flow which, similarly to the conifold theory \cite{KT,KS}, can be interpreted
in terms of a cascade of Seiberg dualities, have a very different (deep) IR physics. Indeed,
these theories generically undergo dynamical supersymmetry breaking (DSB), a pretty novel phenomenon for a 4d field theory embedded in the
context of the gauge/string correspondence.

In this short note we review some of these interesting topics mainly focusing on the $Y^{2,1}$ case. This is a master example as it captures all the interesting physical features of the entire $Y^{p,q}$ class, both in the conformal and in the non conformal cases. More details can be found in
the original papers \cite{MS,noi1,MS2,BHOP,FHSU,noi2}. 

\section{R-charges from holography and $a$-maximization}

There are two interesting AdS/CFT predictions which can be tested, in principle, thanks to supersymmetry and conformal invariance. The first, a direct consequence of the map between the graviton in AdS and the stress energy tensor of the dual field theory, relates the volume of the base of the CY cone, $V(X^5)$, to the gravitational central charge $a$ of the SCFT
\cite{Witten98}
\be
V(X^5) = \frac{~N^2}{4~a}\,\pi^3\ , \, \quad (N>>1)\ .
\label{vola}
\ee
If the exact R-charges of the SCFT are known, the above relation can be explicitly checked, provided the volume of $X^5$ is known\footnote{This is not obviously the case in general. For example, the horizon of the complex cone over the second del Pezzo surface is not explicitly known. Nevertheless the dual field theory can be obtained. Its R-charges were evaluated in \cite{noi1} where old results in the literature were corrected. The value for $a$ thus obtained provided a prediction for the dual volume. Remarkably, this prediction was confirmed once the above volume was evaluated, without any need of the explicit metric, using the geometric dual of the $a$-maximization prescription \cite{msy,bz}.}. In fact, the central charge $a$ of an ${\cal
N}=1$ SCFT can be exactly determined in terms of a certain combination of
linear and cubic 't Hooft anomalies of the $U(1)_R$ current \cite{Anselmi97}\footnote{For SCFT's with matter in 2-index representations (such as quivers), there is no distinction between $a$
and the other gravitational central charge $c$, at leading order in the
large $N$ limit. This is confirmed by holographic
calculations \cite{hs} and is indeed expected to hold for any SCFT admitting a
weakly coupled dual supergravity background (this means that SCFT  with
fundamental matter cannot have a weakly coupled holographic dual since $Tr R$ is not subleading in $1/N$ in this case; see \cite{biga05} for a recent review of the argument).} 
\be
a = \frac{3}{32} \left(3 Tr R^3 - Tr R \right)\ .
\ee
The R-charges of the chiral matter superfields in the SCFT can be holographically predicted thanks to another relevant relation. This is based on the identification of D3-branes wrapped on
supersymmetric three-cycles $\Sigma_A$ with baryonic operators $B_A \sim (X_A)^N$, with $X_A$ being matter chiral primary fields
\cite{Witten298}. In
particular, the R-charges of such baryons are expected to be related to the volumes
of the cycles by
\be
\label{wrabra}
R(B_A)= N~R(X_A) = \frac{\pi}{3} ~ \frac{V(\Sigma_A)}{V(X^5)}~N. 
\ee
All this concerns the holographic predictions. What about the values for $a$ and the R-charges as deduced from a purely field theory analysis? There are two basic conditions the $U(1)_R$ symmetry of an ${\cal N}=1$ SCFT has to satisfy. The first is that it has to be free from ABJ anomalies. Now, the R-symmetry current of the SCFT is in the same superconformal multiplet as the stress energy tensor $T$ and, as a consequence, the above condition is equivalent to the requirement that the gauge coupling beta functions at the IR fixed point are zero. The other condition to be imposed is that if any superpotential $W$ is present at the fixed point, it has to have R-charge equal to two. 
Other conditions are dictated by the symmetries of the theory. For example, it is expected that the $U(1)_R$-symmetry commutes with any non abelian factor of the global non anomalous symmetry group of the SCFT. Moreover, if the latter contains also abelian factors $U(1)_I$ (commuting with the non abelian ones) two classes of non trivial exact relations emerge \cite{iw03}. They involve certain 't Hooft anomalies of the R-current and the symmetry currents $F_I$ of the above $U(1)$'s
\be 
9Tr R^2 F_I = Tr F_I\ ; \quad   Tr RF_IF_J<0\ .
\label{rela}
\ee
These relations, originating\footnote{For instance the first one descends from a proportionality between the anomaly triangles $\langle RRF_I\rangle$ and $\langle TTF_I\rangle$. The exact coefficient can be deduced in the free case where all the fermionic components of the matter superfields have R-charge $-1/3$.} again from the peculiar structure of the superconformal multiplets, contribute to fix an ambiguity, in the definition of the exact R-symmetry, due to the presence of $U(1)_I$ factors in the global symmetry group. 
In this case, in  fact, if $R_0$ is a valid generator of the $U(1)_R$-symmetry (i.e. if it is ABJ-anomaly free, gives charge $2$ to $W$, and commutes with the global non abelian symmetries), so is $R_t = R_0 + \sum_Ix_IF_I$, with $x_I$ apriori arbitrary real numbers. The latter are fixed by imposing the relations (\ref{rela}) or, equivalently, by locally maximizing the ``trial $a$-charge'' $a_t = (3/32)(3 Tr R_t^3 - Tr R_t)$ with respect to $x_I$.  The exact $a$ charge of the theory is then given by the maximal value of $a_t$. Notice that all the abelian anomaly free 
factors (commuting with the non-abelian ones) enter in the extremization process. Hence, if for instance a superpotential breaks some non-abelian global
symmetry and provides extra $U(1)$ factors, these can (and do) mix with the putative R-charge and enter the
$a$-maximization procedure. This is a trivial but crucial observation for the following AdS/CFT checks to work.  

The SCFT duals to the $Y^{p,q}$ are expected to have a global symmetry group which is $SU(2)\times U(1)\times U(1)\times U(1)_R$. Hence the $a$-maximization prescription will be required to fix the exact R-charges of the theory. Moreover it can be expected from the counting of the $U(1)$'s that this will amount on fixing two unknowns. 

\section{The new dual pairs and the AdS/CFT duality checks}
 Sasaki-Einstein manifolds can be divided into three classes according to whether the orbits of a Killing vector, the Reeb vector, are closed (regular SE), have orbifold-like singularities (quasi-regular SE) or are non-compact (irregular SE). This vector is mapped, through the AdS/CFT correspondence, into the R-symmetry current of the dual SCFT. Examples of regular SE manifolds are $S^5$ and $T^{1,1}$. The $Y^{p,q}$ are instead quasi-regular or irregular SE and their volumes
\be
\label{vol5}
V(Y^{p,q}) = \frac{q^2\,\left(2p+\sqrt{4p^2-3q^2}\right)}{3p^2\,
\left(3q^2-2p^2+p\,\sqrt{4p^2-3q^2} \right)}\,\pi^3\ ,
\ee
are respectively rational or irrational multiples of $\pi^3$. Four kinds of three-dimensional supersymmetric submanifolds
$\Sigma_A$ are embedded in the $Y^{p,q}$. Their volumes are $(p,q)$-dependent
fractions (rational/irrational numbers in the quasi-regular/irregular case) of the volume of a unit three-sphere. Four classes of matter fields, each labeled by the value of the R-charge, are then expected, due to the holographic prediction (\ref{wrabra}), in the dual field theories. 
 Identifying the latter is not an easy task. One can gain some insight by studying two simple singular (i.e. topology-changing) limits of the volumes above. When $q\rightarrow p$, $V(Y^{p,q})\rightarrow V(S^5/{\mathbb Z}_{2p})$, while for $q\ll p$, $V(Y^{p,q})\rightarrow V(T^{1,1}/{\mathbb Z}_p)$. The dual gauge theories in these two different limits have both gauge group $SU(N)^{2p}$. One could thus conjecture that this will be the gauge group for any value of $q,p$. This heuristic observation is in fact confirmed by a detailed analysis of the toric geometrical properties of the cones over $Y^{p,q}$ \cite{MS}, which remarkably contain all the relevant informations to extract the whole structure of the dual field theories. 
 These \cite{MS2} contain $4p+2q$ chiral bifundamental matter superfields, $2p+2q$ of which being in $SU(2)$ doublets, 
interacting through a superpotential $W$ which is a sum of $2q$ cubic and $p-q$ quartic $SU(2)$ invariant terms. 
We will not review here the exact structure of the SCFT's in the whole class. Instead we will focus on the irregular $Y^{2,1}$ case which captures, however, all the interesting physics.

It was realized in \cite{MS} that the CY cone over $Y^{2,1}$ is the complex cone over the first del Pezzo surface. The low energy effective dynamics on $N$ D3-branes at the tip of that singularity \cite{hanane} 
%
is an $SU(N)^4$ SCFT which can be represented in terms of the quiver diagram in figure 1.
\begin{figure}[ht]
\begin{center}
{\includegraphics{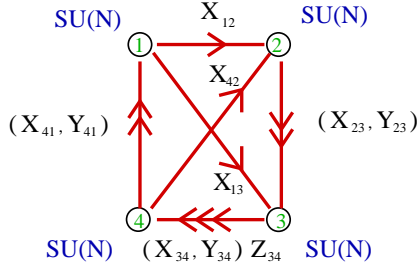}}
\caption{\small Each node here represents a gauge factor. The arrows refer to bi-fundamental chiral multiplets eventually appearing in $SU(2)$ doublets (grouped in parenthesis).}
\label{reg1}
\end{center}
\end{figure}
There is also a superpotential which is the sum of two cubic and one quartic $SU(2)$ invariants terms. It is explicitly given by
\bea
\label{sup}
W&=&\mbox{Tr}\;[X_{34} Y_{41}X_{13} - Y_{34} X_{41}X_{13}- X_{34} X_{42} Y_{23} +
\nonumber \\&+&
Y_{34} X_{42}X_{23} + Z_{34} X_{41} X_{12} Y_{23}- Z_{34} Y_{41} X_{12} X_{23}]\ .
\eea
The R-charges and $a$-charge of this theory were evaluated in the literature, but they did not match with those inherited from the geometry of $Y^{2,1}$. The mismatch was cured in \cite{noi1} by the following simple observation. The non anomalous flavor symmetry group of the theory without superpotential
would be $SU(3)_{34}\times SU(2)_{41}\times SU(2)_{32}\times U(1)\times U(1)$. However $W$ in (\ref{sup}) breaks
it to $SU(2) \times U(1)\times U(1)$ where one of the $U(1)$'s comes in fact from the breaking of the
non-abelian $SU(3)_{34}$ factor. Hence, the global symmetry group of the SCFT is $SU(2)\times U(1)\times U(1)\times U(1)_R$, as expected from holography. Only $3p+q=7$ apriori independent R-charges are left over by the non abelian symmetries. In the literature the symmetry breaking induced by the superpotential was not taken into account and so only $6$ R-charges were taken to be apriori independent. This produced wrong results when $a$-maximization was employed.


There are 4 ($2p$ in the general case) conditions on the R-charges coming from the vanishing of the gauge coupling beta functions. They are all of the form 
\be
T(G) + (1/2)\sum_{f}(R_f-1) + (1/2)\sum_{a}(R_a-1)=0\ ,
\ee
where $G=SU(N), T(G)=N$ for each node of the quiver and the sums extend over the fundamental (f) and 
antifundamental (a) fields coupling with the node. Moreover there are 3 (i.e. $p+q$) conditions from 
$R(W)=2$. Crucially, only 5 ($3p + q -2$ in the general case) of the whole 7 conditions are linearly independent.  
The best one can do by solving these equations is thus to parameterize the R-charges of all the chiral superfields in terms of two unknowns, say $x$ and $y$, naturally related with the two abelian factors in the global non-R symmetry group. This is a very general feature for all the SCFT duals to $Y^{p,q}$. 

The explicit parameterization in the $Y^{2,1}$ case reads 
\bea
\label{sist}
R_{13}=R_{42}=R^{(Z)}_{34} \equiv x,\  R_{14}=R_{23}\equiv y,\ 
R^{(X)}_{34}=R^{(Y)}_{34}=2-x-y,\ R_{12}=2-x-2y.
\eea
The trial $a$-charge $a_t$ in this case can be easily expressed in terms of $x$ and $y$ as 
\be
a_t(x,y) = \frac{9 N^2}{32}[4 + 4 (y-1)^3 + 3 (x-1)^3 +
(1 -x -2y)^3 + 2 (1-x-y)^3]\ .
\ee
This has a maximum at $x_{m} = -3 + \sqrt{13}~,~ y_{m}  = \frac{4}{3}(4 - \sqrt{13})$. Using this result 
one finally gets that the SCFT above has irrational central charge
\be
a = c = a_t(max) \,=\, N^2 \, \left(-46 + 13 \sqrt{13} \right)\ .
\ee
The corresponding irrational R-charges are: $- 3 + \sqrt{13}\ $ for $(Z_{34},\,X_{13},\,X_{42})$, $\frac 13 (-17 + 5 \sqrt{13})\ $ for $X_{12}$, $\frac 43 (4 - \sqrt{13})\ $ for $(X_{23},Y_{23},X_{41},Y_{41})$ and $\frac 13 (-1 + \sqrt{13})\ $ for $(X_{34},Y_{34})$.

One can then construct four (just as the number of supersymmetric three-cycles in the dual geometry) 
classes of color singlet baryons, each class being labeled by the R-charge $R(B_{ij}) = N \,R_{ij}$. 

The irrational volume of $Y^{2,1}$ (eq. (\ref{vol5}) for $p=2, q=1$) as well as those of its three-cycles \cite{MS,HEK} exactly match with the above field theory findings through the holographic relations (\ref{vola}), (\ref{wrabra}).

All what we have derived so far can be (and has been \cite{MS2}) repeated for any $Y^{p,q}$. The story
goes on pretty much the same. In particular, $a$-maximization, which plays a crucial r\^ole in the correspondence, works 
along the same lines since the number of unknowns, from a field theory point
of view, is in all cases equal to two. In other words, although the gauge theory gets more and more complicated
as far as $p$ and $q$ increase, the interesting physics is essentially the same for the whole series.

\section{Breaking conformal invariance}

Let us now consider the addition of $M$ fractional D3-branes to the stack of $N$ D3's at the 
tip of the  cone over  $Y^{2,1}$. This breaks conformal invariance in the gauge theory side and 
the dual supergravity background is expected to be modified (as NSNS and RR 3-fluxes will be induced).

The non conformal gauge theory \cite{FHHW} has gauge group
$SU(N)\times SU(N+3M)\times SU(N+M)\times SU(N+2M)$  (here the factors refer to the nodes 1,2,3,4 
respectively) while the matter content and the superpotential are, in form, as in 
the conformal case. 

The anomalous dimensions of the matter
fields in the non conformal theory approach the ones in the $M=0$
case in the limit $M\ll N$ and, due to the symmetries of the quiver diagram, get corrections only at order
$(M/N)^2$. Using this observation it is easy to deduce that, in the limit, the four gauge coupling 
$\beta$-functions of the deformed theory are proportional to
\bea
&& b_2 = M(10-\sqrt{13}) > 0\ ,\qquad \qquad \, \, b_1=-b_2\ , \nonumber \\
&& b_4 = M(7\sqrt{13}-22) > 0\ , \qquad \qquad b_3=-b_4\ . 
\eea 
As can be seen
from the above expressions, couples of couplings run in opposite
directions. For generic initial conditions the gauge factor with
higher rank (node 2) will run to strong coupling first. Hence, in a given
interval of energies, one can approximate the gauge theory
dynamics by taking the other three nodes as weakly coupled and
acting as flavor symmetries, with the actual dynamics being driven
by the strongly coupled gauge factor. 
In order to follow the flow towards the IR we can apply a Seiberg duality on node 2. 
This way one gets the same quiver structure and the same superpotential as before, with only 
$N\rightarrow N-M$. The fact that the superpotential remains invariant in structure after 
Seiberg duality is not so evident in principle, since it contains both cubic and quartic terms. 
However one can show \cite{noi2} that this is indeed the case.

The new node 2 still reaches strong coupling first. One can thus Seiberg dualize again, and so on. After $k$ duality steps, $N\rightarrow N-kM$. As for the conifold case \cite{KT,KS}, the flow towards the IR can be described in terms of 
a selfsimilar cascade of Seiberg dualities along which the effective number of
 regular D3-branes is reduced by units of $M$. This was shown to happen in general \cite{HEK} for every non conformal deformation of the duals of $Y^{p,1}$ and $Y^{p,p-1}$ SE manifolds.

One expects these features to be reflected by the dual supergravity background.
First, since along the cascade the structure of the non conformal theory differs from the one in the conformal case only in the gauge group, the mesonic branch of the moduli space will be again described by the singular original CY cone. Hence a convenient metric ansatz can be given by a warped product of flat 4d Minkowski and the singular cone, just as is done in the conformal case. Second,
the above results on the beta functions translate \cite{FHHW} in the following predictions for the dilaton and the flux of $B_2$ on the topological $S^2$ 
\be
\sum_{i=1}^4 {8\pi^2\over g_i^2}\approx e^{-\Phi}={\rm const}\ ;\quad  3{8\pi^2\over g_2^2}+{8\pi^2\over g_3^2}+2{8\pi^2\over g_4^2}= M(8+4\sqrt{13})\log r={e^{-\Phi}\over2\pi\alpha'}\int_{S^2}B_2\ .
\ee
Here the adimensional field theory energy ratio $r$ is in fact mapped to the radial coordinate of the dual background, so that for large (small) radius the dual field theory is in the UV (IR). Since the cascade amounts on effective decreasing the number of regular D3's, an $r$ dependent flux for $F_5$ is expected, as well as a constant $F_3$ flux counting the number of fractional branes. 

A solution having these properties and exactly encoding the physics of the cascade has been found in \cite{HEK}. However this is singular at small $r$ just as the singular Klebanov-Tseytlin \cite{KT} solution for the conifold model. There, a regular solution was found \cite{KS} by carefully considering the IR gauge theory physics. In our case a similar care is needed.

In fact, the selfsimilar cascade continues until, at some IR scale $\Lambda$, $N_f \leq N_c$ for some node: when this happens the moduli space gets
quantum corrections, a $\Lambda$-dependent term is introduced into
the superpotential and the theory is not self similar
anymore along the RG flow. To analyze the theory below $\Lambda$ one has to take care of the above corrections and so replace the singular cone in the metric ansatz with some different stuff. This results to be a deformed conifold for the theory in \cite{KS}. The deformation amounts on blowing up the $S^3$ of $T^{1,1}$ so that the $F_3$ flux can be supported in the IR without divergences and the dual IR physics (displaying confinement and chiral symmetry breaking) is precisely encoded.

A natural question to answer is whether something similar happens in our cases. The are geometric arguments \cite{altmann,fhu} which actually forbid supersymmetric complex deformations (with a non shrinking $S^3$) for the CY cones over $Y^{p,q}$, thus suggesting that the dual physics should be pretty
different from the conifold one. In what follows, we try to match these geometric
predictions with the dual IR gauge theory results.
\subsection{The dynamics at the end of the cascade}
If $N=kM$, after $k-1$ cascade steps the gauge group of the non conformal theory in the $Y^{2,1}$ case is reduced to $SU(M)\times
SU(4M)\times SU(2M)\times SU(3M)$ (in the general cases where the cascade proceeds selfsimilarly it reduces to $SU(M)\times SU(2M)\dots\times SU(2pM)$). This  is where the moduli space
gets modified since the node with higher rank has now an equal number of colors and
flavors ($N_f = N_c = 4M$, or $2pM$ in the general case).  This quantum modification can be
taken care by suitably deforming the superpotential as 
\bea
\label{supb} 
W&=&\mbox{Tr}\;[X_{34} Y_{41}X_{13} - Y_{34}
X_{41}X_{13}- X_{34} {\cal M}^Y_{43}+ Y_{34}{\cal M}^X_{43} +
\nonumber \\&+& Z_{34} X_{41}{\cal M}^Y_{13}-Z_{34} Y_{41}{\cal
M}^X_{13}] +  \xi (\det{\cal M}-B{\tilde B}-\Lambda^{8M})\ ,  
\eea
where $\xi$ is a Lagrange multiplier and the meson matrix ${\cal
M}$ is formed by the bilinears
\bea 
{\cal M}^X_{13}=X_{12}X_{23} ~,~ \quad {\cal
M}^X_{43}=X_{42}X_{23} ~,~ \quad {\cal M}^Y_{13}=X_{12}Y_{23} ~,~ \quad {\cal
M}^Y_{43}=X_{42}Y_{23}\ , 
\eea 
while the baryons are \bea B=
(X_{23})^{2M}(Y_{23})^{2M}\ , \qquad {\tilde B}=
(X_{12})^{M}(X_{42})^{3M}\ . 
\eea 
Flavor and gauge indexes are understood in the above expressions.
 
The F-term equations for $\xi, B$ and ${\tilde B}$ have two
possible classes of solutions. In the mesonic branch $\ B={\tilde B}=0\ ,\ \det {\cal
M}=\Lambda^{8M}$. The F-term equations with respect to $Y_{34}$
and $X_{34}$ give ${\cal M}^X_{43}= X_{41}X_{13}$ and ${\cal
M}^Y_{43}= Y_{41}X_{13}$ which, after fixing the D-term equations
for the other nodes, implies ${\cal M}^X_{43}=0$, hence finally
$\det {\cal M}=0$ running to an inconsistency. The mesonic branch is therefore empty.

The baryonic branch is $\xi=\det {\cal M}=0 ~,~ B{\tilde
B}=-\Lambda^{8M}\ $. As it is evident from the structure of the
superpotential (\ref{supb}), the mesons ${\cal M}^X_{43}$ and
${\cal M}^Y_{43}$ get mass and can be integrated out. Therefore,
at energy scales below $\Lambda$, the dynamics of node 2 decouples
from the rest and the theory reduces to the quiver triangle in figure 2. 
\begin{figure}[ht]
\begin{center}
{\includegraphics{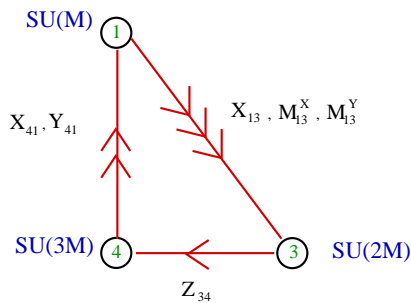}}
\caption{\small The quiver triangle at the bottom of the cascade.}
\end{center}
\end{figure}
\noindent 
Now is node 4 to reach strong coupling first, and the meson reads ${\cal N}=
\left( {\cal N}^X_{31}~,~{\cal N}^Y_{31}\right)\equiv\left(Z_{34}
X_{41}~,~Z_{34} Y_{41}\right)$. 
Since for this node $N_f (=2M) <
N_c (=3M)$, the superpotential includes an Affleck-Dine-Seiberg
non-perturbative contribution  and reads 
\bea 
W=\mbox{Tr}\;[{\cal
N}^X_{31}{\cal M}^Y_{13}-{\cal N}^Y_{31}{\cal M}^X_{13}]+ M
\left({\Lambda^{7M}\over \det {\cal N}}\right)^{{1\over M}}\ . 
\eea
This superpotential admits no supersymmetric vacua. Indeed, the F-term
equations for ${\cal M}_{13}$ give 
\be {\cal N}_{31}=0\ , \ee 
whereas those for ${\cal N}_{31}$ give 
\be {\cal
M}_{13}-\frac{\Lambda^{7}}{( \det {\cal N})^{\frac1M +1}}{\rm
Minor}({\cal N}_{31})=0\ , 
\ee 
the two solutions being inconsistent.

The conclusion is that supersymmetry is dynamically broken
(similar results have been reached in \cite{BHOP,FHSU}). Notice that this is an
infrared effect driven by the dynamically generated superpotential term.
If the dual smooth background exists, it has to be a
non-supersymmetric deformation of the solution in \cite{HEK}. This agrees (although only qualitatively) with the geometric
expectations, which indeed forbid supersymmetric complex
deformations for the original CY cones.

Notice that while the logic is similar to the conifold case, the
physics is different here, as we see. In the former case the
baryonic branch has solutions and indeed the theory is at some
point on this branch: the strongly coupled gauge group decouples
below the corresponding strong interacting scale, leaving a non
interacting massless goldstone mode while the IR dynamics is
driven by the left over $SU(M)$ factor. The theory hence confines, chiral symmetry is broken down to
${\mathbb Z}_2$, and supersymmetry is preserved. In the present case the
strongly interacting node decouples, too, but the left over gauge
theory is described by the quiver triangle and 
does not admit any supersymmetric vacua.

The question whether the  non supersymmetric 
background exists and is stable or unstable could be answered by a careful study 
of the non supersymmetric vacua of the theory. This is a program which was started 
in \cite{BHOP,FHSU} and which will be important to further develop. 
The possibility of having a non supersymmetric background with
known field theory dual is of evident interest. For instance, once embedded
into a compact CY, this could provide a concrete realization of
D-brane cosmological scenarios where supersymmetry
breaking would not be put by hand, but
actually would arise spontaneously, as a dynamical effect encoded
in the theory itself. 

\vskip 5pt 
\centerline{\bf Acknowledgments} 
\vskip 5pt \noindent
Work partially  supported by the European
Commission RTN Program MRTN-CT-2004-005104, MRTN-CT-2004-503369,
CYT FPA 2004-04582-C02-01, CIRIT GC 2001SGR-00065 and by MIUR.
M.B. is also supported by a MIUR fellowship within the program
``Incentivazione alla mobilit\`a di studiosi stranieri e italiani
residenti all'estero''.



\begin{thebibliography}{99}
{\small 
\bibitem{malda}
J.~M.~Maldacena,
 Adv.\ Theor.\ Math.\ Phys.\  {\bf 2} (1998) 231
[Int.\ J.\ Theor.\ Phys.\  {\bf 38} (1999) 1113], hep-th/9711200.
%
\bibitem{dimers} A. Hanany, K. D. Kennaway, {\it Dimer models and toric diagrams}, hep-th/0503149. S. Franco, A. Hanany, K. D. Kennaway, D. Vegh, B. Wecht, {\it Brane dimers and quiver gauge theories}, hep-th/0504110.
%
\bibitem{vafa} B. Feng, Y. He, K. D. Kennaway, C. Vafa, {\it Dimer models from mirror symmetry and quivering amoebae}, hep-th/0511287.
%
\bibitem{kw}
I. R. Klebanov, E. Witten,
Nucl. Phys. B {\bf 536} (1998) 199, hep-th/9807080.
%
\bibitem{gmswSE}
J. P. Gauntlett, D. Martelli, J. Sparks, D. Waldram,
Adv.\ Theor.\ Math.\ Phys. {\bf 8} (2004) 711, hep-th/0403002.
 
%
\bibitem{kn} I. R. Klebanov, N. A. Nekrasov, Nucl. Phys.  B {\bf 574} (2000) 263, hep-th/9911096.
%
\bibitem{iw03}
K. Intriligator, B. Wecht, 
Nucl. Phys. B {\bf 667} (2003) 183, hep-th/0304128.
%
\bibitem{KT} I.R. Klebanov, A.A. Tseytlin,
Nucl. Phys. B {\bf 578} (2000) 123, hep-th/0002159.
%
\bibitem{KS} I.R. Klebanov, M. Strassler,
JHEP {\bf 08} (2000) 52, hep-th/0007191.
%
\bibitem{MS} D. Martelli, J. Sparks,
{\it Toric Geometry, Sasaki--Einstein Manifolds and a new Infinite
Class of AdS/CFT duals}, hep-th/0411238.
%
\bibitem{noi1}
M.~Bertolini, F.~Bigazzi, A.L.~Cotrone, 
JHEP {\bf 12} (2004) 024, hep-th/0411249.
%
\bibitem{MS2}
S.~Benvenuti, S.~Franco, A.~Hanany, D.~Martelli, J.~Sparks,
JHEP {\bf 0506} (2005) 064, hep-th/0411264.
%
\bibitem{BHOP} D. Berenstein, C.P. Herzog, P. Ouyang, S. Pinansky,
JHEP {\bf 0509} (2005) 084, hep-th/0505029.
%
\bibitem{FHSU}  S. Franco, A. Hanany, F. Saad, A.M. Uranga,
{\it Fractional Branes and Dynamical Supersymmetry Breaking}, 
hep-th/0505040.
%
\bibitem{noi2}
M.~Bertolini, F.~Bigazzi, A.L.~Cotrone, 
Phys. Rev. D {\bf 72} (2005) 061902, hep-th/0505055.
%
%
\bibitem{Witten98}
S.S. Gubser, I.R. Klebanov, A.M. Polyakov,
Phys. Lett. B {\bf 428} (1998) 105, hep-th/9802109.
E.~Witten,
Adv. Theor. Math. Phys. {\bf 2} (1998) 253, hep-th/9802150.
%
\bibitem{msy}D. Martelli, J. Sparks, S.T. Yau, {\it The Geometric Dual of a-maximisation for Toric Sasaki-Einstein Manifolds}, hep-th/0503183.
%
\bibitem{bz} A. Butti, A. Zaffaroni, {\it R-charges from toric diagrams and the equivalence of a-maximization and Z-minimization}, hep-th/0506232. 
%
\bibitem{Anselmi97}
D. Anselmi, D.Z. Freedman, M.T. Grisaru, A.A. Johansen,
Nucl. Phys. B {\bf 526} (1998) 543, hep-th/9708042.
%
\bibitem{hs}
M. Henningson, K. Skenderis,
JHEP {\bf 07} (1998) 023, hep-th/9806087.
%
\bibitem{biga05}
F. Bigazzi, R. Casero, A.L. Cotrone, E. Kiritsis, A. Paredes,
JHEP {\bf 0510} (2005) 012,
hep-th/0505140.
%
\bibitem{Witten298}
E. Witten, 
JHEP {\bf 07} (1998) 6, hep-th/9805112.
D. Berenstein, C.P. Herzog, I.K. Klebanov,
JHEP {\bf 06} (2002) 047, hep-th/0202150.
%
\bibitem{hanane}
B.~Feng, A.~Hanany, Y.~H.~He,
Nucl.\ Phys.\ B {\bf
595} (2001) 165, hep-th/0003085.
%
\bibitem{FHHW}
S. Franco, Y. H. He, C. Herzog, J. Walcher,
Phys. Rev. D {\bf 70} (2004) 046006, hep-th/0402120.
%
\bibitem{HEK} C. P. Herzog, Q. J. Ejaz, I. R. Klebanov, 
JHEP {\bf 0502} (2005) 009,
 hep-th/0412193.
%
\bibitem{altmann}
K. Altmann, {\it The versal deformation of an isolated toric Gorenstein singularity},
alg-geom/9403004.
%
\bibitem{fhu}  S. Franco, A. Hanany, A, M. Uranga, JHEP 0509 (2005) 028, hep-th/0502113.
}
\end{thebibliography}
\end{document}